\documentclass[a4paper]{jpconf}
\usepackage{graphicx}
\begin{document}
\title{Strictly localised triplet dimers on one- and two-dimensional lattices}

\author{S. Jackson and J.H.Samson }

\address{Department of Physics, Loughborough University, Loughborough, LE11 3TU, UK}

\ead{s.jackson3@lboro.ac.uk, j.h.samson@lboro.ac.uk}

\begin{abstract}
Electrons may form inter-site pairs (dimers) by a number of mechanisms. For example, long-range (Fr\"{o}hlich) electron-phonon interactions and strong on-site Hubbard U allow formation of small light bipolarons in some lattices.  We identify circumstances under which triplet dimers are strictly localised by interference in certain one- and two-dimensional lattices. We assume a U-V Hamiltonian with nearest- and next-nearest-neighbour hopping integrals $t$ and $t^{'}$, large positive $U$ and attractive nearest- and next-nearest-neighbour interactions $V$ and $V^{'}$.  In the square ladder and some two-dimensional bilayers, if the dimer Hilbert space is restricted to nearest- and next-nearest-neighbour dimers, triplet dimers become strictly localised for certain values of these parameters. For example, in a square ladder with $t^{'} = t$ and $V^{'} = V$, all triplet bands become flat due to exact cancellation of hopping paths.  We identify the localised eigenstates for all flat bands in each lattice. We show that many of the flat bands persist for arbitrary $t/ t^{'}$ so long as other restrictions still apply.
\end{abstract}


Electrons may form inter-site pairs (dimers) by a number of mechanisms. For example, long-range (Fr\"{o}hlich) electron-phonon interactions and strong on-site Hubbard U allow formation of small light bipolarons in some lattices.  We identify circumstances under which triplet dimers are strictly localised by interference in certain one- and two-dimensional lattices.  
Our calculations are not specific to the pairing mechanism but our model can be obtained from the Coulomb-Fr\"{o}hlich model \cite{Anisotropic,Frohlich,AK99,Spencer, superlight 1, superlight 2}. 
In this model electrons interact with lattice distortions on neighbouring sites. The electron together with the distortion forms a new quasiparticle -- the polaron. An effective attraction between polarons can exist because two polarons can deform the lattice more effectively together than separately.


Applying the Lang-Firsov transformation \cite{Lang} we obtain the $UV$ model, 
valid for strong coupling \cite{AM}:
\begin{equation}\label{uv}
H=-\sum_{ ij\sigma} \ t_{ij} c^{\dagger}_{i\sigma}\ c_{j\sigma} + U\sum_{i}\ n_{i\uparrow}\ n_{i\downarrow}
+\frac{1}{2}\sum_{i}\sum_{j\neq i}\sum_{\sigma \sigma^{'}}V_{ij} c^{\dagger}_{i\sigma} c_{i\sigma}c^{\dagger}_{j\sigma^{'}} c_{j\sigma^{'}}
\end{equation}
where $ c^{\dagger}_{i\sigma}$ and $ c_{j\sigma} $ are electron creation and annihilation operators respectively, $t_{ij}$ is a renormalized hopping,
$U$ is the onsite Coulomb repulsion 
and $V_{ij}$ is the polaron-polaron potential containing the Coulomb repulsion between electrons on neighbouring sites $i$ and $j$ and the (non-retarded) effective attraction due to the Fr\"{o}hlich electron-phonon interaction (EPI). 
We take the on-site repulsion $U$ to be infinite so that basis states with two electrons on a single site are suppressed.

The Hilbert space of two particles on on a $d$-dimensional tight-binding lattice can be represented as a particle on a $2d$-dimensional lattice, the tensor product of the two Hilbert spaces.  However, if the attraction between the particles is strong enough to bind them into a dimer over the whole Brillouin zone, the low-lying states will have large amplitude only near a $d$-dimensional subspace where the particles are in close proximity.  Accordingly a truncation of the Hilbert space of dimers to a small number of bond lengths will capture the essential physics.  We call the truncated Hilbert space of the dimers the \emph{dimer lattice}.
Let 
\begin{equation}
\label{Di}
D_{i}=\{j:0<r_{ij} \le L_\mathrm{max}\}
\end{equation}
be the set of sites $j$ whose distance $r_{ij}$ from a site $i$ is no more than $L_\mathrm{max}$. We shall call  $D_{i}$ the neighbours of $i$.  
A dimer will have (spin-independent) diagonal potential $V_{ij}$.  We write $V$ for the nearest-neighbour potential and $V^{'}$ for the next-nearest-neighbour potential.

If a lattice $\Lambda$ has $N$ sites and the mean number of neighbours $|D_{i}|$ is $\nu$ then the single-electron Hilbert space is $2N$-dimensional, the two-electron Hilbert space is $N(2N-1)$-dimensional and the dimer Hilbert space is $4 \frac\nu2 N$-dimensional. We can further reduce to one singlet and three triplet spaces, each of dimensionality $\nu N/2$. The dimer spaces for $S_{z}=0$ are

\begin{equation} \label{Hs}
\mathcal{S}=\mbox{span }  \left\{ \frac{1}{\sqrt{2}}  \left( \left|i\uparrow j\downarrow \right\rangle+\left|j\uparrow i\downarrow \right\rangle\right): i\in \Lambda, j\in D_{i} \right\}  
\end{equation}

\begin{equation}\label{Ht0}
\mathcal{T}_{0}=\mbox{span }  \left\{ \frac{1}{\sqrt{2}}  \left( \left|i\uparrow j\downarrow \right\rangle-\left|j\uparrow i\downarrow \right\rangle\right): i\in \Lambda, j\in D_{i} \right\}  
\end{equation}\label{zero}

for singlets and triplets respectively. In each case basis states are double-counted in the span.  We note that triplets are spatially antisymmetric and singlets are symmetric; if we represent the above dimer basis states as arrows pointing from $i$ to $j$, then $|i\longrightarrow j\rangle = \pm|j\longrightarrow i\rangle$ with $+$ for singlets. The above formalism enables us to write the Hamiltonian of the dimers in each sector as a tight-binding Hamiltonian on a dimer lattice constructed by placing a node on the line joining each site $i$ to each
point $j\in D_{i}$. If $j\in D_{i}$ and $k\in D_{i}$, and $t_{jk}\neq 0$, then the dimer can hop from $ij$ to $ik$. A dimer hopping vector is then drawn between the two nodes on the dimer lattice with hopping integral $t_{jk}$. 


%

\begin{figure}[h]
	\centering
		\includegraphics[width=4.5in]{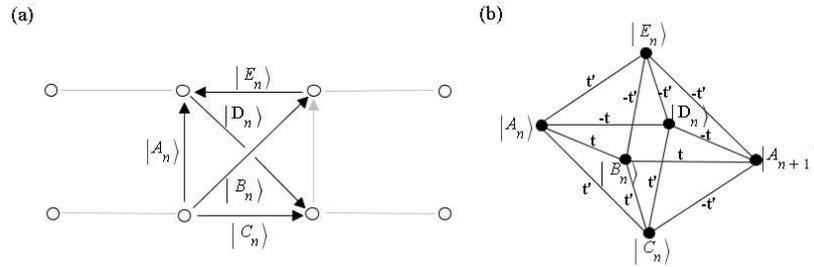}
\caption{(a) Triplet basis states in the square ladder for nearest- and next-nearest-neighbour dimers. (b) Unit cell of the triplet dimer lattice.  Adjacent cells are connected by the points $\left| A,n \right\rangle $ which form bottlenecks.}
	\label{basis}
\end{figure}

We consider a square ladder with nearest- and next-nearest-neighbour dimers, nearest and next-nearest-neighbour hopping $t$ and $t^{'}$ respectively and nearest- and next-nearest-neighbour interactions $V$ and $V^{'}$ respectively.
Each sector of the truncated Hilbert space may be reduced to the five basis states indicated in figure \ref{basis}(a), where $n$ is the index of the unit cell.
The dimer lattice comprises a chain of corner-sharing octahedra as indicated in figure \ref{basis}(b).
We find that some or all of the bands are flat depending on the ratios $V^{'}/V$ and $t^{'}/t$.
For $V^{'}=V$ we obtain the triplet band structure: 
\begin{equation}
E(k)-V=\pm 2t
\end{equation}
\begin{equation}
E(k)-V=0
\end{equation}
\begin{equation}
E(k)-V=\pm  \sqrt{ 8t^{2}\cos^{2}{ \left( \frac{ka}{2}\right) } +8t'^{2}\sin^{2} { \left( \frac{ka}{2}\right)} } 
\end{equation}
For $V^{'}=V$ and $t^{'}=t$ we obtain the triplet band structure $E(k)-V= 0$, $E(k)-V=\pm 2t$, $E(k)-V=\pm2 \sqrt{2}t$. 
Since all bands are flat we infer that triplets are strictly localised for $V^{'}=V$ and $t^{'}=t$.
This is because the points $\left| A,n \right\rangle $ in the dimer lattice (see figure \ref{basis}(b)) are bottlenecks meaning that a dimer propagating along the ladder must repeatedly pass through these points.
Any path from $\left| A,n \right\rangle $ to a neighbouring bottleneck $\left| A,n \pm 1 \right\rangle $ can be replaced by a path of opposite sign (and equal magnitude if $t^{'}=t$) by interchanging $\left| B,n \right\rangle $ with $\left| C,n \right\rangle $ and $\left| D,n \right\rangle $ with $\left| E,n \right\rangle $. 
For example $\left| A,n\right\rangle \rightarrow \left| B,n\right\rangle \rightarrow \left| A,n+1\right\rangle$ is replaced with $\left| A,n\right\rangle \rightarrow \left| C,n\right\rangle \rightarrow -\left| A,n+1\right\rangle$.
Since each path can be replaced by another path of opposite sign, and equal magnitude if $t^{'}=t$, the paths cancel each other for $t^{'}=t$ and triplet dimers become strictly localised. 
We have identified five linearly-independent localised eigenstates of the Hamiltonian.

We also consider square and honeycomb bilayers with restricted dimer-lengths and find that if some additional restrictions apply all triplet bands are flat for $V^{'}=V$ and $t^{'}=t$. As in the square ladder this is due to the existence of bottlenecks and the exact cancellation of paths connecting neighbouring bottlenecks. 
The band structure and localised eigenstates for these bilayers are essentially the same as for the square ladder. 

We conclude that triplets are less mobile than singlets on a square ladder and that for $t^{'}=t$ and $V^{'}=V$, if dimer-lengths are restricted, triplets become strictly localised. This is also true for square and honeycomb bilayers. 



\end{document}